\documentclass[runningheads]{llncs}
\usepackage[T1]{fontenc}
\usepackage{graphicx}
\usepackage{xurl}
\usepackage{hyperref}
\usepackage{amsmath}
\usepackage{booktabs}
\usepackage{bbding}
\usepackage{tabularx}
\begin{document}
\begin{sloppypar}
\title{UltraUPConvNet: A UPerNet- and ConvNeXt-Based Multi-Task Network for Ultrasound Tissue Segmentation and Disease Prediction}
\titlerunning{UltraUPConvNet for Ultrasound}
%
%
\author{Zhi Chen\inst{1}\and
Le Zhang\inst{1}}
\authorrunning{Z. Chen, L. Zhang}
%
\institute{School of Engineering, College of Engineering and Physical Sciences, University of Birmingham, Birmingham, UK\\
\email{zxc561@student.bham.ac.uk}}
\maketitle              
\begin{abstract}

Ultrasound imaging is widely used in clinical practice due to its cost-effectiveness, mobility, and safety. However, current AI research often treats disease prediction and tissue segmentation as two separate tasks and their model requires substantial computational overhead. In such a situation, we introduce UltraUPConvNet, a computationally efficient universal framework designed for both ultrasound image classification and segmentation. Trained on a large-scale dataset containing more than 9,700 annotations across seven different anatomical regions, our model achieves state-of-the-art performance on certain datasets with lower computational overhead. Our model weights and codes are available at \url{https://github.com/yyxl123/UltraUPConvNet}.

\end{abstract}
\section{Introduction}

Ultrasound imaging is widely adopted in clinical practice due to its cost-effectiveness, mobility, and safety.  As deep learning models have shown strong performance in applications such as computer vision and natural language processing, deep learning models for ultrasound have also shown great potential. Recently, General Medical Artificial Intelligence (GMAI) models \cite{zhang2023challengesperspectivesfoundationmodels} have demonstrated robust generalization capabilities across a wide range of tasks and datasets. Although some GMAI models have been proposed, they typically require substantial computational overhead and have relatively complex architectures, highlighting the need for a model specifically for ultrasound with lower computational overhead and simpler architectures. 

Recent advances in deep learning models for ultrasound imaging have addressed some specific tasks, such as breast tumor prediction \cite{floyd1994prediction}. However, when faced with new tasks or datasets, they overlook intrinsic dataset task relationships and need extensive retraining. The focus is now on universal models, known for their versatility in segmentation tasks across domains. At the same time, some universal models like MedSAM \cite{ma2024segment}, SAM-Med2D \cite{cheng2023sammed2d}, UniUSNet \cite{lin2024uniusnet} and SAMUS \cite{lin2023samus} are characterized by the heavy stacking of Transformer blocks, leading to substantial computational costs and increased architectural complexity.

Most methods perform well in segmentation while struggling with classification.
But our ConvNeXt-UperNet-based network can do classification task
more effectively. For ultrasound’s complexity, we propose
fully automated prompts that encode prior knowledge.

Therefore, we propose UltraUPConvNet, a universal and promptable framework
for ultrasound that addresses multiple clinical tasks. Our contributions include: 
\begin{itemize}
    \item Versatile Model Framework: Our framework uses four model prompts to obtain better performance and flexibility in various clinical tasks. 
    \item    Efficient and Simple Model Architecture: Our framework does not adopt the currently popular Transformer architecture, but instead relies entirely on convolutional operations, resulting in lower computational complexity and a simpler model structure.
    \item Extensive Experiments: The model outperforms single dataset and ablated versions, achieving state-of-the-art performance and strong generalization to new domains.
\end{itemize}

\section{Method}
\subsection{Multi-task Paradigm}
Multi-task learning is a machine learning method that enhances the generalization ability of each task by simultaneously learning multiple related tasks within the same model. In our specific task, the proposed framework is required to perform both disease prediction and tissue segmentation, which correspond to a classification task and a segmentation task, respectively.

Therefore, the proposed model incorporates two dedicated decoders, one for classification and the other for segmentation. The overall loss is computed in an alternating fashion, where segmentation and classification batches are processed separately during each training epoch. This strategy avoids task interference and maintains task-specific supervision, while still benefiting from a shared feature extractor.

\begin{figure}
\centering 
\includegraphics[width=\textwidth]{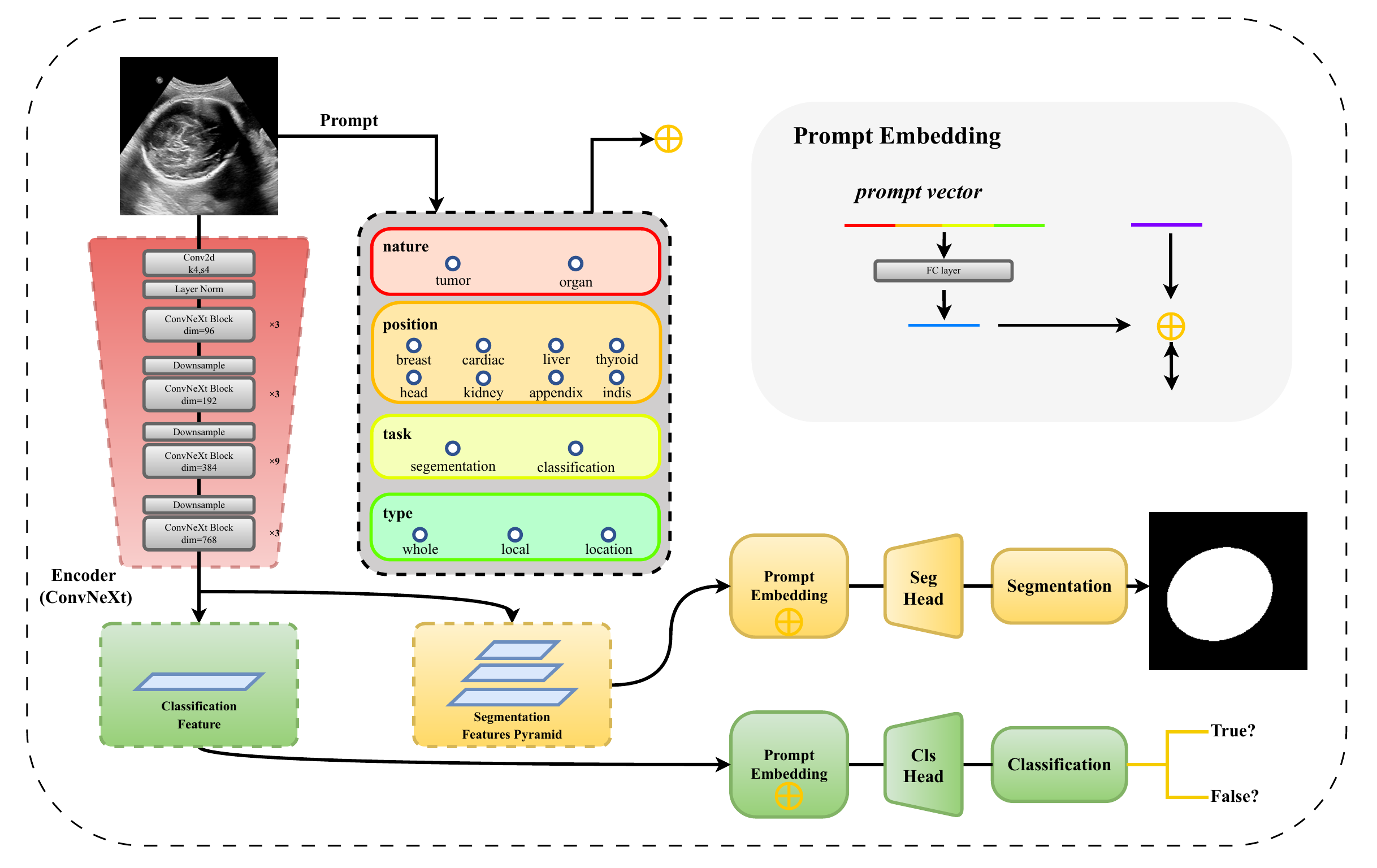}
\caption{The architecture of the proposed UltraUPConvNet.} \label{fig1}
\end{figure}

\subsection{Model Architectures}
UltraUPConvNet (Fig. \ref{fig1}) is a general encoder-decoder
model. It also uses prompts to simultaneously help with
multiple ultrasound tasks like segmentation and classification.
The encoder extracts features, while task-specific decoders are
enhanced by four types of prompts—nature, position, task, and
type—added to extracted features via prompt projection
embedding, boosting the model’s versatility and performance.

\subsubsection{Encoder} Inspired by UniUSNet's good performance, we would like to try a better and simpler backbone without using Transformer. 
We finally chose ConvNeXt \cite{9879745}. ConvNeXt is an efficient convolutional neural network (CNN) model whose design integrates the advantages of traditional convolutional neural networks and Transformer models, aiming to provide more efficient computational performance while maintaining high accuracy. 
To make our model lightweight, we chose ConvNeXt-Tiny, which is the most compact variant in the ConvNeXt family while still maintaining competitive performance.

\subsubsection{Prompting Strategy} In our specific task, four prompts—nature, position, task, and type—are used to provide information about images and tasks. This method can possibly enhance the model’s flexibility and interpretability. Learning from Vision Transformers (ViT) \cite{dosovitskiy2020image} and \cite{lin2024uniusnet}, our work defines four prompt one-hot vectors and project them into specific dimensions using fully connected (FC) layers in both classification and segmentation tasks.

\subsubsection{Decoder} Since we want to have a lightweight model, we choose UperNet \cite{xiao2018unified} as our segmentation decoder.  UperNet is built upon the Feature Pyramid Network (FPN) and PPM (Pyramid Pooling Module). It is proven to excel at semantic segmentation of medical images \cite{zhang2025medical}. Specifically, FPN combines high-resolution low-level features with semantically rich high-level features through a top-down pathway and lateral connections. To match the input dimensionality, we also modified the dimensions of UperNet accordingly.

\subsection{Loss Function}

\subsubsection{Classification} 
For the segmentation task, we adopt a compound loss that balances between pixel-wise classification and region-wise consistency. Specifically, we use a weighted combination of Cross Entropy Loss and Dice Loss. Given the model's output $x_{\text{seg}}$ and the ground truth label $y$, the total segmentation loss is defined as:
\[
\mathcal{L}_{\text{seg}} = 0.4 \cdot \mathcal{L}_{\text{CE}}(x_{\text{seg}}, y) + 0.6 \cdot \mathcal{L}_{\text{Dice}}(x_{\text{seg}}, y)
\]
Here, $\mathcal{L}_{\text{CE}}$ is the standard cross-entropy loss, while $\mathcal{L}_{\text{Dice}}$ is the soft Dice loss computed over the softmax outputs of the segmentation map. This combination enhances both per-pixel accuracy and overall region segmentation quality.

\subsubsection{Segmentation} 
The classification task is designed to handle both binary (2-way) and multi-class (4-way) scenarios within a single training loop. During each iteration, the number of classes required for each sample is indicated by a batch-level metadata flag. Accordingly, two separate classifier heads are employed: one for 2-way classification and another for 4-way classification. The total classification loss is computed as:
\[
\mathcal{L}_{\text{cls}} = 
\begin{cases}
\mathcal{L}_{\text{CE-2}}(x_{\text{cls-2}}, y) & \text{if 2-way} \\
\mathcal{L}_{\text{CE-4}}(x_{\text{cls-4}}, y) & \text{if 4-way}
\end{cases}
\]
where $x_{\text{cls-2}}$ and $x_{\text{cls-4}}$ denote the classification outputs from the respective classifier heads. If a batch contains both types of tasks, the final loss is the sum of the two. This design allows for flexible handling of mixed-task datasets.

\subsubsection{Overall Loss Strategy}
The overall loss is computed in an alternating fashion, where segmentation and classification batches are processed separately within each training epoch. Specifically, segmentation loss and classification loss are computed from their respective data loaders in separate loops. This strategy enables task-specific optimization while allowing the model to learn shared representations across tasks.

To balance the gradient contributions from both tasks, the classification loss is scaled by a weighting coefficient $\lambda_{\text{cls}}$ during backpropagation. The final loss for backpropagation is defined as:
\[
\mathcal{L}_{\text{final}} = 
\begin{cases}
\lambda_{\text{seg}} \cdot \mathcal{L}_{\text{seg}}, & \text{for segmentation batches} \\
\mathcal{L}_{\text{cls}}, & \text{for classification batches}
\end{cases}
\]

This weighted design helps stabilize multi-task training by preventing the classification task from dominating or under-contributing to the optimization process. In our experiments, $\lambda_{\text{cls}}$ is set to 10, empirically based on validation performance.

\section{Experimental Results}
\subsection{Implementation Details} 
The dataset, as shown in Table. ~\ref{tab1}, was split into training,
validation, and testing sets (7:1:2). We also used ConvNeXt's official pre-trained weights. Learning from \cite{lin2024uniusnet} , we trained our
models for 200 epochs using AdamW (initial learning rate: 2e-5) with data augmentation techniques like random flipping,
rotation (-20° to 20°), and cropping. Since our model is highly lightweight, it can be trained using an RTX 2060 with only 6GB of VRAM. Implementation was done
in PyTorch.


\begin{table}
\centering 
\caption{DATASET DESCRIPTION.}\label{tab1}
\begin{tabular}{|c|c|c|c|c|}
\hline
Dataset &Public?&  Position & Image Num & Annotation\\
\hline
BUSI \cite{al2020dataset} &\Checkmark& Breast & 780 & Classification, Segmentation \\
BUSIS \cite{zhang2022busis}&\Checkmark& Breast & 562 & Segmentation \\
BUS-BRA \cite{gomez2024bus} &\Checkmark& Breast & 1875 & Classification, Segmentation \\
Fatty-Liver \cite{byra2018transfer} &\Checkmark& Liver & 550 & Classification \\
kidneyUS \cite{singla2023open}&\Checkmark& Kidney & 534 & Segmentation \\
DDTI \cite{pedraza2015open} &\Checkmark& Thyroid & 466 & Segmentation \\
Fetal HC \cite{van2018automated} &\Checkmark& Head & 999 & Segmentation \\
CAMUS  \cite{leclerc2019deep}&\Checkmark& Cardiac & 500 & Segmentation \\
Appendix \cite{marcinkevivcs2023regensburg}&\Checkmark& Appendix & 474 & Classification \\
Appendix &\XSolidBrush & Appendix & 67 & Classification \\
Breast &\XSolidBrush & Breast & 150 & Classification, Segmentation \\
Breast-luminal &\XSolidBrush & Breast & 236 & Classification, Segmentation \\
Cardiac &\XSolidBrush & Cardiac & 76 & Segmentation \\
Fetal Head &\XSolidBrush & Head & 121 & Segmentation \\
Kidney &\XSolidBrush & Kidney & 67 & Segmentation \\
Liver &\XSolidBrush & Liver & 103 & Classification \\
Thyroid  &\XSolidBrush & Thyroid & 428 & Segmentation \\
\hline
\end{tabular}
\end{table}

\subsection{Results}


\begin{table}[ht]
\centering
\caption{OVERALL PERFORMANCE COMPARISON ON SEVERAL DATASETS.}
\label{tab2}
\resizebox{\textwidth}{!}{%
\begin{tabular}{c c c c c}
\toprule
\textbf{Dataset} & SAMUS & UniUSNet & UltraUPConvNet w/o prompt & \textbf{UltraUPConvNet} \\
& (Interactive) & (Automatic) & (Automatic) & (Automatic)\\
\midrule
\multicolumn{1}{c}{\textbf{Params}} & 130.10M & 86.29M & 60.44M& 60.48M\\
\midrule
BUS-BRA   & 83.88\% & 75.59\% & 88.37\%& \textbf{88.46\%}\\
BUSIS     & 88.89\% & 87.84\% &\textbf{92.23\%}& 91.33\%\\
CAMUS     & 72.44\% & 92.58\% & 94.62\%& \textbf{94.71\%}\\
DDTI      & 69.70\% & 66.06\% & 79.25\%& \textbf{80.55\%}\\
Fetal\_HC & 97.60\% & 96.69\% & \textbf{97.72\%}& 97.11\%\\
KidneyUS  & 67.54\% & \textbf{96.03\%}& 90.02\%& 89.49\%\\
\midrule
\textbf{Seg Average} & 80.01\% & 85.80\% & \textbf{90.37\%}& 90.28\%\\
\midrule
Appendix   & / & 52.84\% & 75.89\%&\textbf{77.30\%}\\
BUS-BRA    & / & 80.39\%&  90.96\%& \textbf{92.02\%} \\
Fatty-Liver& / & 89.36\%&  \textbf{100\%} & \textbf{100.00\%} \\
\midrule
\textbf{Cls Average} & / & 74.20\% & 88.95\%& \textbf{89.77\%} \\
\midrule
\textbf{Total Average} & / & 81.93\% & 89.90\% & \textbf{90.11\%} \\
\bottomrule
\end{tabular}
}
\end{table}

\subsubsection{Comparison with SOTA}
We compared our model with state-of-the-are (SOTA) models, like SAMUS and UniUSNet.
SAMUS is a SAM variant for ultrasound data and UniUSNet is basd on a modified Swin-Unet. From \cite{lin2024uniusnet}, we get SAMUS's and UniUSNet's 
results trained on BroadUS-9.7K Dataset.
From Table. ~\ref{tab2}, SAMUS's segmentation performance(80.01\%) is lower than UniUSNet(85.80\%), while SAMUS has better segmentation performance on more datasets.
Compared to UniUSNet, our automatic prompt model, with 29.9\% fewer parameters, achieves better both 
segmentation(90.28\%) and classification(89.95\%)
results. 

\begin{figure}
\centering 
\includegraphics[width=0.7\textwidth]{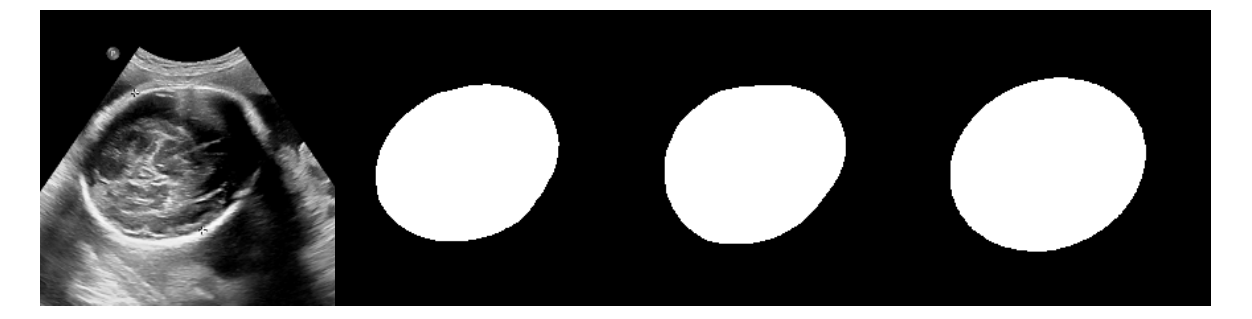}
\caption{Some examples of segmentation result. From left to
right: original image, UltraUPConvNet w/o prompt, UltraUPConvNet
and ground truth.} \label{fig2}
\end{figure}

\subsubsection{Ablation Study}
We also trained our UltraUPConvNet model
without prompts to assess whether prompts can enhance
performance. The segmentation results in Fig.~\ref{fig2} reveal that UltraUPConvNet performs well on the segmentation task.
From Table. ~\ref{tab2}, ablation study reveals that UltraUPConvNet
(90.11\%) outperforms the ablation version (89.90\%), proving
the effectiveness of prompts. 

\section{Discussion and Conclusion}
We propose a lightweight universal model for medical ultrasound imaging
that integrates four prompts. Experiments on some public datasets
validate its effectiveness. We will further assess the model’s adaptability in our future work.

\begin{credits}

\subsubsection{\ackname} 
This work was conducted in participation of the Universal UltraSound Image Challenge (MICCAI 2025, Daejeon, Republic of Korea), focusing on multi-organ classification and segmentation.

\end{credits}
%
%
%
\bibliographystyle{splncs04}
\bibliography{refs}

@inproceedings{lin2024uniusnet,
  title={UniUSNet: A Promptable Framework for Universal Ultrasound Disease Prediction and Tissue Segmentation},
  author={Lin, Zehui and Zhang, Zhuoneng and Hu, Xindi and Gao, Zhifan and Yang, Xin and Sun, Yue and Ni, Dong and Tan, Tao},
  booktitle={2024 IEEE International Conference on Bioinformatics and Biomedicine (BIBM)},
  pages={5550--5557},
  year={2024},
  organization={IEEE}
}

@INPROCEEDINGS{9879745,
  author={Liu, Zhuang and Mao, Hanzi and Wu, Chao-Yuan and Feichtenhofer, Christoph and Darrell, Trevor and Xie, Saining},
  booktitle={2022 IEEE/CVF Conference on Computer Vision and Pattern Recognition (CVPR)}, 
  title={A ConvNet for the 2020s}, 
  year={2022},
  volume={},
  number={},
  pages={11966-11976},
  doi={10.1109/CVPR52688.2022.01167}}

@misc{zhang2023challengesperspectivesfoundationmodels,
      title={On the Challenges and Perspectives of Foundation Models for Medical Image Analysis}, 
      author={Shaoting Zhang and Dimitris Metaxas},
      year={2023},
      eprint={2306.05705},
      archivePrefix={arXiv},
      primaryClass={eess.IV},
      url={https://arxiv.org/abs/2306.05705}, 
}

@article{ma2024segment,
  title={Segment anything in medical images},
  author={Ma, Jun and He, Yuting and Li, Feifei and Han, Lin and You, Chenyu and Wang, Bo},
  journal={Nature Communications},
  volume={15},
  number={1},
  pages={654},
  year={2024},
  publisher={Nature Publishing Group UK London}
}

@misc{cheng2023sammed2d,
      title={SAM-Med2D}, 
      author={Junlong Cheng and Jin Ye and Zhongying Deng and Jianpin Chen and Tianbin Li and Haoyu Wang and Yanzhou Su and Ziyan Huang and Jilong Chen and Lei Jiang and Hui Sun and Junjun He and Shaoting Zhang and Min Zhu and Yu Qiao},
      year={2023},
      eprint={2308.16184},
      archivePrefix={arXiv},
      primaryClass={cs.CV},
      url={https://arxiv.org/abs/2308.16184}, 
}

@article{lin2023samus,
  title={Samus: Adapting segment anything model for clinically-friendly and generalizable ultrasound image segmentation},
  author={Lin, Xian and Xiang, Yangyang and Zhang, Li and Yang, Xin and Yan, Zengqiang and Yu, Li},
  journal={arXiv preprint arXiv:2309.06824},
  volume={4},
  number={11},
  year={2023},
  publisher={Sep}
}

@inproceedings{xiao2018unified,
  title={Unified perceptual parsing for scene understanding},
  author={Xiao, Tete and Liu, Yingcheng and Zhou, Bolei and Jiang, Yuning and Sun, Jian},
  booktitle={Proceedings of the European conference on computer vision (ECCV)},
  pages={418--434},
  year={2018}
}

@article{zhang2025medical,
  title={Medical image segmentation by combining feature enhancement Swin Transformer and UperNet},
  author={Zhang, Lin and Yin, Xiaochun and Liu, Xuqi and Liu, Zengguang},
  journal={Scientific Reports},
  volume={15},
  number={1},
  pages={14565},
  year={2025},
  publisher={Nature Publishing Group UK London}
}

@article{dosovitskiy2020image,
  title={An image is worth 16x16 words: Transformers for image recognition at scale},
  author={Dosovitskiy, Alexey and Beyer, Lucas and Kolesnikov, Alexander and Weissenborn, Dirk and Zhai, Xiaohua and Unterthiner, Thomas and Dehghani, Mostafa and Minderer, Matthias and Heigold, Georg and Gelly, Sylvain and others},
  journal={arXiv preprint arXiv:2010.11929},
  year={2020}
}

@article{al2020dataset,
  title={Dataset of breast ultrasound images},
  author={Al-Dhabyani, Walid and Gomaa, Mohammed and Khaled, Hussien and Fahmy, Aly},
  journal={Data in brief},
  volume={28},
  pages={104863},
  year={2020},
  publisher={Elsevier}
}

@inproceedings{zhang2022busis,
  title={BUSIS: a benchmark for breast ultrasound image segmentation},
  author={Zhang, Yingtao and Xian, Min and Cheng, Heng-Da and Shareef, Bryar and Ding, Jianrui and Xu, Fei and Huang, Kuan and Zhang, Boyu and Ning, Chunping and Wang, Ying},
  booktitle={Healthcare},
  volume={10},
  number={4},
  pages={729},
  year={2022},
  organization={MDPI}
}

@article{gomez2024bus,
  title={BUS-BRA: a breast ultrasound dataset for assessing computer-aided diagnosis systems},
  author={G{\'o}mez-Flores, Wilfrido and Gregorio-Calas, Maria Julia and Coelho de Albuquerque Pereira, Wagner},
  journal={Medical Physics},
  volume={51},
  number={4},
  pages={3110--3123},
  year={2024},
  publisher={Wiley Online Library}
}

@article{byra2018transfer,
  title={Transfer learning with deep convolutional neural network for liver steatosis assessment in ultrasound images},
  author={Byra, Micha{\l} and Styczynski, Grzegorz and Szmigielski, Cezary and Kalinowski, Piotr and Micha{\l}owski, {\L}ukasz and Paluszkiewicz, Rafa{\l} and Ziarkiewicz-Wr{\'o}blewska, Bogna and Zieniewicz, Krzysztof and Sobieraj, Piotr and Nowicki, Andrzej},
  journal={International journal of computer assisted radiology and surgery},
  volume={13},
  number={12},
  pages={1895--1903},
  year={2018},
  publisher={Springer}
}

@inproceedings{singla2023open,
  title={The open kidney ultrasound data set},
  author={Singla, Rohit and Ringstrom, Cailin and Hu, Grace and Lessoway, Victoria and Reid, Janice and Nguan, Christopher and Rohling, Robert},
  booktitle={International Workshop on Advances in Simplifying Medical Ultrasound},
  pages={155--164},
  year={2023},
  organization={Springer}
}

@inproceedings{pedraza2015open,
  title={An open access thyroid ultrasound image database},
  author={Pedraza, Lina and Vargas, Carlos and Narv{\'a}ez, Fabi{\'a}n and Dur{\'a}n, Oscar and Mu{\~n}oz, Emma and Romero, Eduardo},
  booktitle={10th International symposium on medical information processing and analysis},
  volume={9287},
  pages={188--193},
  year={2015},
  organization={SPIE}
}

@article{van2018automated,
  title={Automated measurement of fetal head circumference using 2D ultrasound images},
  author={van den Heuvel, Thomas LA and de Bruijn, Dagmar and de Korte, Chris L and Ginneken, Bram van},
  journal={PloS one},
  volume={13},
  number={8},
  pages={e0200412},
  year={2018},
  publisher={Public Library of Science San Francisco, CA USA}
}

@article{leclerc2019deep,
  title={Deep learning for segmentation using an open large-scale dataset in 2D echocardiography},
  author={Leclerc, Sarah and Smistad, Erik and Pedrosa, Joao and {\O}stvik, Andreas and Cervenansky, Frederic and Espinosa, Florian and Espeland, Torvald and Berg, Erik Andreas Rye and Jodoin, Pierre-Marc and Grenier, Thomas and others},
  journal={IEEE transactions on medical imaging},
  volume={38},
  number={9},
  pages={2198--2210},
  year={2019},
  publisher={IEEE}
}

@article{marcinkevivcs2023regensburg,
  title={Regensburg pediatric appendicitis dataset},
  author={Marcinkevi{\v{c}}s, Ri{\v{c}}ards and Reis Wolfertstetter, Patricia and Klimiene, Ugne and Chin-Cheong, Kieran and Paschke, Alyssia and Zerres, Julia and Denzinger, Markus and Niederberger, David and Wellmann, Sven and Ozkan, Ece and others},
  journal={(No Title)},
  year={2023},
  publisher={Zenodo}
}

@article{floyd1994prediction,
  title={Prediction of breast cancer malignancy using an artificial neural network},
  author={Floyd Jr, Carey E and Lo, Joseph Y and Yun, A Joon and Sullivan, Daniel C and Kornguth, Phyllis J},
  journal={Cancer: Interdisciplinary International Journal of the American Cancer Society},
  volume={74},
  number={11},
  pages={2944--2948},
  year={1994},
  publisher={Wiley Online Library}
}





\end{sloppypar}
\end{document}